\documentclass[aps,prd, nofootinbib]{revtex4}

\usepackage{graphicx}
\usepackage{dcolumn}
\usepackage{amssymb}
\usepackage{hyperref}
\usepackage[all]{xy}
\usepackage[dvips]{epsfig}
\usepackage{graphicx}
\usepackage{amsfonts,amsmath}
\DeclareGraphicsExtensions{.ps,.eps,.pdf,.png}

\usepackage[usenames,dvipsnames]{xcolor}

\newcommand{\half}{\frac{1}{2}}
\newcommand{\LCDM}{$\Lambda$CDM}

\begin{document}

\title{Cosmological effects of coupled dark matter}

\author{Sophie C. F. Morris}
\email{ppxsm@nottingham.ac.uk}
\author{Anne M. Green}
\email{anne.green@nottingham.ac.uk}
\author{Antonio Padilla}
\email{antonio.padilla@nottingham.ac.uk}
\author{Ewan R. M. Tarrant}
\email{ppxet@nottingham.ac.uk}

\affiliation{School of Physics and Astronomy, 
University of Nottingham, Nottingham, NG7 2RD, UK}

\begin{abstract}
Many models have been studied that contain more than one species of dark matter and some of these couple the Cold Dark Matter (CDM) to a light scalar field. In doing this we introduce additional long range forces, which in turn can significantly affect our estimates of cosmological parameters if not properly accounted for. It is, therefore, important to study these models and their resulting cosmological implications. We present a model in which a fraction of the total cold dark matter density is coupled to a scalar field. We study the background and perturbation evolution and calculate the resulting Cosmic Microwave Background anisotropy spectra. The greater the fraction of dark matter coupled to the scalar field and the stronger the coupling strength, the greater the deviation of the background evolution from \LCDM. Previous work, with a single coupled dark matter species, has found an upper limit on the coupling strength of order $\mathcal{O}(0.1)$. We find that with a coupling of this magnitude more than half the dark matter can be coupled to a scalar field without producing any significant deviations from \LCDM.

\end{abstract}

\maketitle

\section{Introduction} \label{sec:intro}

Since the first discovery that visible matter does not account for all the mass in the Universe~\cite{Zwicky:1933gu} astronomers have known that there is something missing from our basic understanding of the Universe. The two most popular explanations for this missing mass have been that there is some other matter present that we cannot see, in the form of Cold Dark Matter (CDM), or that our theory of gravity is inadequate, giving rise to a plethora of modified theories of gravity (for a review see e.g.~Ref.~\cite{Clifton:2011jh}). More recently, many more accurate measurements of the cosmological parameters have been made e.g.~Refs.~\cite{Ade:2013lta,Hinshaw:2012fq,Seljak:2006bg,Tegmark:2006az}, constraining the amount of CDM in the universe to be roughly 22\% of the total energy budget. Cosmological data analyses typically assume a \LCDM\ model in which the accelerated expansion is due to a small cosmological constant and gravity is described by General Relativity, universally coupled to all species of matter (visible and dark). Any deviation from these assumptions could affect the best fit cosmological parameters (see e.g.~Ref.~\cite{Jennings:2009qg}). In this paper, we will alter the latter assumption, allowing for violations of the Equivalence Principle in the dark sector. In particular, we will study a generalised Brans--Dicke scenario in which some species of dark matter couple to the Brans--Dicke scalar, while others do not. \LCDM\ can be identified with the limit in which the coupling vanishes, or else the limit in which the fraction of dark matter coupled to the Brans--Dicke field goes to zero. Here we will explore what happens as the coupling and the fraction of coupled dark matter are increased.

Since the exact nature of dark matter is unknown, it is important to study different models and their effects upon cosmological observables. In particular, plenty of recent work has focussed on the case where dark matter is coupled to a quintessence scalar field playing the role of dark energy e.g.~Refs.~\cite{Ratra:1987rm,Wetterich:1994bg,Martin:2008qp}. Because the quintessence field is necessarily light, this will mediate an additional long range force on dark matter, breaking the gravitational equivalence with baryonic matter.  The astrophysical and cosmological implications of an additional long range force acting exclusively on dark matter were studied  long ago by Gradwohl and Frieman~\cite{Friedman:1991dj, Gradwohl:1992ue,Frieman:1993fv}, and later by Gubser and Peebles~\cite{Gubser:2004uh,Gubser:2004du}, and others~\cite{Nusser:2004qu,Kesden:2006zb,Kesden:2006vz}.  Prior to that, effects on  time varying Newton's constant were used to place bounds on a generalised Brans--Dicke scenario in which visible and dark matter both couple to the Brans--Dicke field, but with differing strengths~\cite{PhysRevLett.64.123}.

If one assumes more than one species of dark matter,  as suggested by the richness of string compactifications~\cite{Chialva:2012rq,Grana:2005jc}, one can also entertain the possibility that the Equivalence Principle might be violated within the dark matter sector itself. Indeed, there is no good theoretical reason to maintain it once we have abandoned the universal coupling between matter and gravity by allowing dark matter to feel an additional long range force.  This possibility was explored to some extent in Ref.~\cite{Brookfield:2007au}, where different dark matter species are coupled to a scalar field with different strengths. In that work the scalar field acts under the influence of a quintessence potential, playing the role of dark energy. This gives rise to chameleon--like behaviour~\cite{Khoury:2003aq,Khoury:2003rn}, with the scalar  settling in to a minimum of the effective potential controlled by the density of the coupled fields. In contrast, here we assume that dark energy is a cosmological constant, and that there is no bare potential for the scalar -- in other words, it is exactly massless. This alters the picture considerably as there is no longer a local environmental minimum for the effective potential. Such a scenario is easily achieved in higher dimensional compactifications. For example, in the two--brane Randall--Sundrum model~\cite{Randall:1999ee}, the warped extra dimension gives rise to exactly the sort of scenario we are interested in, with an exactly massless scalar (the radion) coupled with different strength to matter on different branes~\cite{Garriga:1999yh}.

If they are not properly accounted for, additional long range forces in the dark sector can dramatically affect our estimates for cosmological parameters. This point was emphasized in Ref.~\cite{Kaloper:2009nc},  where ignorance of the additional long range force was shown to result in an erroneously inferred equation of state for dark energy~\footnote{In Ref.~\cite{Kaloper:2009nc} the additional long range force is mediated by a gauge field and yields an additional repulsive force. Here we are focussing on a scalar mediator which will give rise to an additional attractive force, but it is clear that the principles discussed in Ref.~\cite{Kaloper:2009nc} will still apply, although the effects will work in the opposite direction.}.  We can illustrate the principle by considering the orbit of a planet around a star. One could compute the mass of the star by measuring the acceleration of the planet, and then use Newton's law of gravitation to infer the mass. However, if there were an additional long range force acting between the star and the planet, and we were ignorant of this, it is clear that we would infer the wrong value for the star's mass.  For cosmology, the lesson we learn is that we should ask whether or not  the relevant observational probe is sensitive to the additional long range force. When this only acts on (a fraction of) dark matter, the supernova probe will be directly sensitive to the force as supernovae reside in galaxies, which are dominated by dark matter. Galaxies therefore follow accelerated rather than geodesic trajectories and this ought to be properly accounted for when analysing supernova data~\cite{Kaloper:2009nc}. In this paper, we will focus on a more indirect probe of the scalar force -- the Cosmic Microwave Background (CMB). CMB photons are part of the visible sector, so they do not feel the extra force directly. However, the CMB temperature fluctuations are sensitive to density fluctuations, which are affected by any additional long range forces acting on any form of matter, be it visible or dark.

The paper is structured as follows: in Sec. \ref{setupsec} we describe the setup for our model, then in Sec. \ref{backgroundsec} we study the background evolution and examine the extent to which it deviates from \LCDM. In Sec. \ref{pertsec} we study the evolution of perturbations in our system and the effect the coupling has on CMB anisotropies. In addition to comparing to standard \LCDM, we discuss the physics behind the deviations and find that the main physical effects come from the increased density of CDM at early times, a result that is only indirectly related to the coupling.  Finally, we conclude in Sec. \ref{sec:conc}.

\section{Setup}
\label{setupsec}

Let us begin by describing our setup. We consider a generalised Brans--Dicke theory with a cosmological constant responsible for late time acceleration of the universe. The matter content is divided into two components, one of which consists of baryons, radiation and some fraction of dark matter, and the other which is made up of the remaining fraction of dark matter  coupled to the massless scalar field. The action can be written in the Einstein frame  as follows (see also Ref.~\cite{PhysRevLett.64.123}) 

\begin{equation}
\label{action}
S=\int {\rm d}^4 x \sqrt{-g}\left[\frac{1}{16\pi G}(R-2\Lambda)-\frac{1}{2}(\nabla\phi)^2\right] +  S_{\rm SM}[g_{\mu\nu},...] + S_{\rm c}[g_{\mu\nu},...] + S_{q}[\hat g_{\mu\nu},\psi_{q}]\,,
\end{equation}
where $g$ is the determinant of the metric $g_{\mu\nu}$, $R$ is the corresponding Ricci scalar, $\nabla$ is the corresponding covariant derivative and $\phi$ is the scalar field mediating the additional long range force. The action $S_{\rm SM}[g_{\mu\nu},...]$ contains baryons, photons and massless neutrinos and $S_{\rm c}[g_{\mu\nu},...]$ contains uncoupled CDM. The ellipses denote the standard model fields in $S_{\rm SM}$ and uncoupled dark matter field in $S_{\rm c}$. None  of these fields couple directly to the scalar. The action $S_{q}[\hat g_{\mu\nu},\psi_{q}]$  describes dark matter fields that do couple directly to the scalar. Indeed, coupled dark matter particles, $\delta \psi_q$, follow geodesics of the conformally related metric, 
\begin{equation}
\label{coupling}
\hat g_{\mu\nu}=e^{2\alpha\phi}g_{\mu\nu}  \,,
\end{equation}
where  $\alpha$ is the dimensionful coupling constant, which, without loss of generality, we assume to be positive. The field equations describing the system are given by 
\begin{eqnarray}
&& G^\mu_\nu+\Lambda \delta^\nu_\mu = 8\pi G\left(T_{(\rm SM)\nu}^\mu + T_{({\rm c})\nu}^\mu +T_{(\phi)\nu}^\mu+ T_{(q)\nu}^\mu\right)\,, \\
&&\Box \phi=-\alpha T_{(q)\mu}^\mu \,,
\end{eqnarray}
where $\Box=g^{\mu\nu} \nabla_\mu\nabla_\nu$ is the covariant d'Alembertian, 
\begin{equation}
 T_{(\rm SM)\nu}^\mu=-\frac{2}{\sqrt{-g}} g^{\nu\alpha}\frac{\delta S_{\rm SM}}{\delta  g^{\mu\alpha}}\,, \quad\quad T_{({\rm c})\nu}^\mu=-\frac{2}{\sqrt{-g}} g^{\nu\alpha}\frac{\delta S_{\rm c}}{\delta  g^{\mu\alpha}} \,,
\end{equation}
are the energy-momentum tensors for standard model fields and uncoupled CDM, and
\begin{equation}
T_{(\phi)\mu }^\nu=\nabla^\nu \phi \nabla_\mu \phi-\half \delta^\nu_\mu (\nabla \phi)^2 \,,
\end{equation}
is the energy-momentum tensor for the scalar. In addition,  we define 
\begin{equation}
T_{(q)\nu}^\mu=e^{4\alpha\phi}\hat T_{(q)\nu}^\mu \,,
\end{equation}
where the energy-momentum tensor for the  coupled fields is covariantly conserved  in the conformally related Jordan frame, and is given by
\begin{equation}
\hat T_{(q)\nu}^\mu=-\frac{2}{\sqrt{-\hat g}} \hat g^{\nu\alpha}\frac{\delta S_{q}}{\delta \hat g^{\mu\alpha}} \,.
\end{equation}
Note that $T_{(q)\nu}^\mu$ is not covariantly conserved in the Einstein frame, but satisfies $\nabla_\mu  T_{(q)\nu}^\mu=\alpha T_{(q)\mu}^\mu \nabla_\nu \phi $.

\section{Background Cosmology}
\label{backgroundsec}

To leading order, we consider a homogeneous and isotropic universe described by a flat, Friedmann--Robertson--Walker metric,  ${\rm d}s^2=-{\rm d}t^2+a^2(t) {\rm d}\vec x^2$, and a homogeneous scalar $\phi=\phi(t)$.  Each matter field contributes a perfect fluid
\begin{align}
T_{(\rm SM)\nu}^\mu&=\text{diag} (-\rho_{\rm SM}(t), p_{\rm SM}(t), p_{\rm SM}(t), p_{\rm SM}(t)),\qquad T_{({\rm c})\nu}^\mu=\text{diag} (-\rho_{c}(t), 0, 0, 0) \, \nonumber \\
\hat T_{(q)\nu}^\mu&=\text{diag} (-\hat \rho_q(t),  0, 0, 0)
\end{align}
where $\rho_{\rm SM}=\rho_{\rm b}+\rho_{\rm r} $ and $p_{\rm SM}=p_{\rm b}+p_{\rm r}$, where ${\rm b}$ denotes the baryons and ${\rm r}$ collectively denotes radiation and neutrinos. We have that $p_{\rm b}=0$, $p_{\rm r}=\rho_r/3$ and  $p_\phi=\rho_\phi= \dot \phi^2/2$.  
Now, neither  $\hat T_{(q)\nu}^\mu$ nor $ T_{(q)\nu}^\mu$ is covariantly conserved in the Einstein frame, and by association, neither $\hat \rho_q$ nor $\rho_q=e^{4\alpha \phi} \hat \rho_q$ satisfy the usual conservation law for matter (i.e.  ${\rm d} (a^3 \hat \rho_q )/{\rm d} t \neq 0$,  ${\rm d} (a^3  \rho_q )/{\rm d}t \neq 0$).  However, we can use the fact that  $\hat T_{(q)\nu}^\mu$ is covariantly conserved in the Jordan frame to show that 
$
\rho_* = e^{3\alpha\phi}\hat\rho_{q}=e^{-\alpha \phi} \rho_q   \,
$
scales like  $1/a^3$~\cite{Khoury:2003aq,Khoury:2003rn}. The background evolution equations are given by the Friedmann equation, 
\begin{equation}
H^2 = \frac{8\pi G}{3}\left(\rho_\Lambda+\rho_{\rm SM} +\rho_{\rm c} +\frac{1}{2}\dot\phi^2 + \rho_* e^{\alpha\phi}\right)\,,
\label{Friedmann}
\end{equation}
and the homogeneous equation of motion for the scalar
\begin{equation}
\ddot\phi+3H\dot\phi +\alpha\rho_* e^{\alpha\phi}=0\,,
\label{KGeqn}
\end{equation}
where the Hubble parameter $H=\dot a/a$ and overdots represent differentiation with respect to cosmic time $t$. Note that $\rho_\Lambda=\Lambda/8\pi G$ is constant, while $\rho_{\rm b} \propto \rho_{\rm c} \propto \rho_* \propto 1/a^3$, and $\rho_{\rm r} \propto 1/a^4$.  It is also convenient to define the following density fractions
$$
\Omega_\Lambda=\frac{8\pi G\rho_\Lambda}{3H^2},\qquad \Omega_n=\frac{8\pi G\rho_n}{3H^2}, \qquad \Omega_\phi=\frac{4\pi G\dot \phi^2}{3H^2},\qquad  \Omega_q=\frac{8\pi Ge^{\alpha \phi}\rho_*}{3H^2}, \qquad  \Omega_{\rm r}=\frac{8\pi G\rho_{\rm r}}{3H^2}
$$
where $n$ denotes either baryons or uncoupled CDM, and in analogy with quintessence models, we have identified $\Omega_{\rm DE}=\Omega_\Lambda+\Omega_\phi$. We note that this grouping of dark components is somewhat arbitrary~\cite{Kunz2007Dark}.

Let us  now consider the resulting evolution of the cosmological background. The coupling to (some) dark matter endows the scalar with an effective potential $V_{\rm eff}(\phi)=\rho_* e^{\alpha\phi}$. 

Unlike the standard chameleon scenarios~\cite{Khoury:2003aq,Khoury:2003rn} where a local minimum exists at finite $\phi$ (due to the existence of a bare potential for $\phi$), our effective potential is of run--away form, with a minimum only at $+\infty$. The only way the scalar can be held approximately fixed is by Hubble friction, and only when\footnote{This can be seen by dropping the acceleration term in Eq.~(\ref{KGeqn}), and rearranging to give $\alpha^2\rho_*/3H\approx {\rm d}(e^{-\alpha\phi})/{\rm d} t$. This derivative is then small if $H\gg \alpha^2\rho_*$.}$ \, H\gg \alpha^2\rho_*$. Therefore it is difficult to imagine that the scalar field does not significantly affect the background evolution (unless $\alpha \ll 1/M_{\rm pl}$) in contrast to the assumptions made in Ref.~\cite{Gradwohl:1992ue} where the background is always taken to be equivalent to  \LCDM. 
\begin{figure}[t]
\begin{center}
\includegraphics[width=1.0\textwidth]{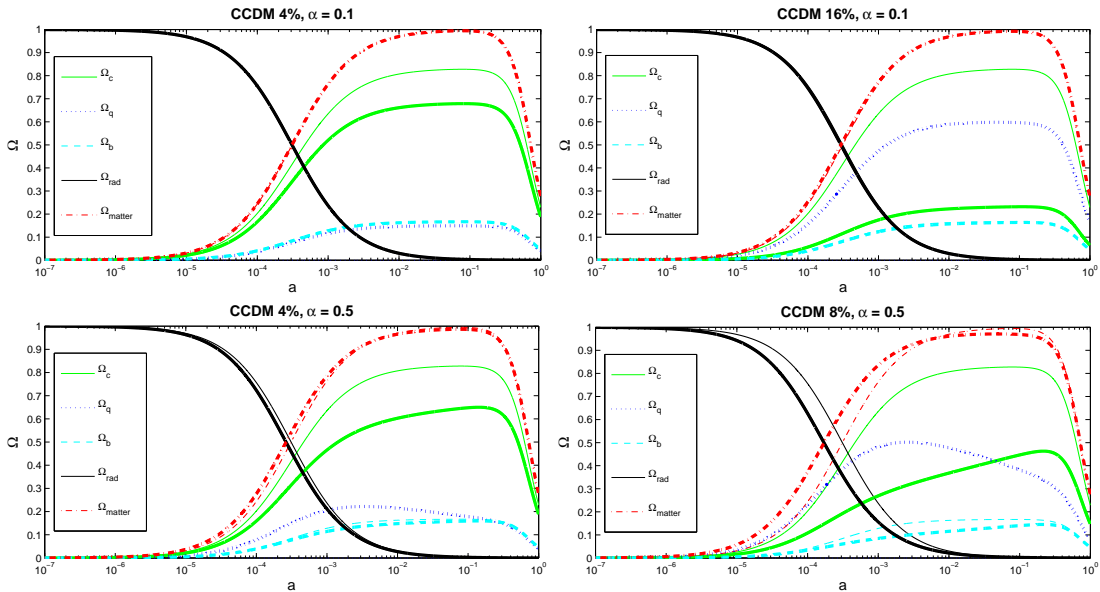}
\caption{The evolution of the density parameters in our model  (thick lines) compared with $\Lambda$CDM  (thin lines), for varying coupling strengths, $\alpha$, and varying proportions of coupled dark matter (CCDM) such that CCDM=4\% means that 4\% of the total energy density in the universe is made up of coupled dark matter.  The energy density of coupled dark matter is
$\Omega_{q}$ and $\Omega_{\rm c}$ is the energy density of uncoupled dark matter. The energy densities have been normalised to their present day values i.e.~the system was evolved to give the same energy densities today as in \LCDM. The coupling constant $\alpha$ is in units of $1/M_{\rm pl}$, where $M_{\rm pl}$ is the Planck mass.}
\label{background}
\end{center}
\end{figure}

These expectations are born out by our numerical solutions plotted in Fig.~\ref{background}.  Here we show the background evolution for various fractions of coupled dark matter today  and various coupling strengths, $\alpha$. We compare these to the corresponding evolution in a \LCDM\ universe, assuming \LCDM\ best--fit cosmological parameter values. We normalise each coupled dark matter model such that they share the same baryon, total dark matter, and dark energy density fraction today ($\Omega^0_{\rm b}$, $\Omega^0_{\rm{c_{tot}}}$, $\Omega_{\rm DE}^0$) as the best fit \LCDM\ model. In the limit $\alpha=0$, we recover \LCDM\ where $\Omega_{\rm DE}^0=\Omega_\Lambda^0$ and $\Omega^0_{\rm{c_{tot}}}=\Omega^0_{\rm c}$. This normalisation is achieved by performing a binary search over the initial condition $\phi_i$ and over the cosmological constant density $\rho_\Lambda$ to find the correct initial conditions necessary in order to realise the desired background. We set $\dot{\phi}_i=0$ since we begin the evolution deep inside the radiation dominated epoch, where we should expect the field to be frozen by Hubble friction. If the coupling is large enough, we can easily see the effect of the evolving scalar field as we move back in time. Initially, at low redshift,  the main effect is to decrease the density fraction stored in baryons (and dark matter), relative to \LCDM.  This is to be expected because the mutual coupling to gravity allows energy to be exchanged between the coupled dark matter and the kinetic energy of the evolving scalar. At higher redshifts, Hubble friction suppresses the kinetic energy of the scalar, but we still see the effect of the scalar's previous evolution on the relative density of radiation to matter. In particular, matter--radiation equality happens at higher redshift in  our coupled dark matter scenario. This is because we need more matter at high redshift to compensate for the relative dip in the matter density fraction at lower redshift, if all cosmological parameters are normalised by their values today. This effect also manifests itself through a deficit in the amount of radiation at high redshift, compared with \LCDM. In any event, it is clear that the background evolution is sensitive to coupling (some) matter to a light scalar. This difference in background evolution is often neglected in the literature (see e.g. Ref.~\cite{Gradwohl:1992ue}).

To further analyse the background dynamics it is instructive to  transform our system into autonomous form  and investigate the presence of any late time attractor solutions. We use the standard definitions~\cite{Copeland:1997et} along with an extra term to include the conformal factor,
\begin{align}
x&\equiv \frac{\dot\phi}{\sqrt{6}H}  \,, &
y&\equiv \frac{1}{H}\sqrt{\frac{\rho_{\gamma}}{3}} \,, \nonumber \\
z&\equiv \frac{1}{H}\sqrt{\frac{\rho_*e^{\alpha\phi}}{3}}\,,&
s&\equiv \frac{1}{H}\sqrt{\frac{\rho_{\Lambda}}{3}} \,.
\end{align}
The flatness condition gives the constraint
\begin{equation}
\Omega_{{\rm b}+{\rm c}}=1-x^2-y^2-z^2-s^2 ,
\end{equation}
where we have grouped baryons and the uncoupled CDM together as they both scale like $a^{-3}$. In these equations $\rho_{\gamma}$ and $\rho_{\Lambda}$ are the radiation and cosmological constant energy densities respectively. Differentiating these equations with respect to $N=\ln a$, which we denote using $^\prime$,  yields the following set of equations

\begin{align} 
x'&= -\left(\frac{H'}{H}+3\right) x - \frac{3}{\sqrt{6}}\alpha z^2  \,, \nonumber \\
y' &= -2y - \frac{H'}{H}y \,, \nonumber \\
z' &= - \left(\frac{H'}{H}+\frac{3}{2}\right)z + \frac{\sqrt{6}}{2}\alpha z x  \,,\nonumber \\
s' &= -\frac{H'}{H}s \,,
\label{autoeqns}
\end{align}
where 
\begin{equation}
\frac{H'}{H}=-\frac{1}{2} (3x^2+y^2-3s^2+3) \,.
\end{equation}
The system (Eqs.~\ref{autoeqns}) generically admits 6 critical points at real values, listed in Table \ref{critpts}.
\begin{table}[h]
\centering
\begin{tabular}{|l|c|c|c|c|}
 \hline
 & $x$ &$y$ & $z$ &$s$ \\
 \hline  \hline
Uncoupled matter domination &0&0&0&0 \\
  \hline
Kinetic Scalar domination & 1 &0 & 0 & 0 \\
  \hline
 Radiation domination & 0 & 1 &0 &0 \\
  \hline
 $\Lambda$ domination & 0 & 0 & 0 & 1 \\
  \hline
 Coupled DM -- Kinetic Scalar scaling& $-\sqrt{\frac{2}{3}} \alpha$ & 0 & $\sqrt{1-\frac{2}{3} \alpha^2}$ & 0 \\
  \hline
 Coupled DM -- Kinetic Scalar -- Radiation scaling & $-\frac{1}{\sqrt{6} \alpha}$ & $\sqrt{1-\frac{1}{2\alpha^2}}$ & $\frac{1}{\sqrt{3} \alpha}$ & 0 \\
  \hline \hline
\end{tabular}
\caption{Critical points for the autonomous system (Eqs.~\ref{autoeqns}).}
\label{critpts}
\end{table}

\begin{figure}[h]
\begin{center}
\includegraphics[width=150mm,height=90mm]{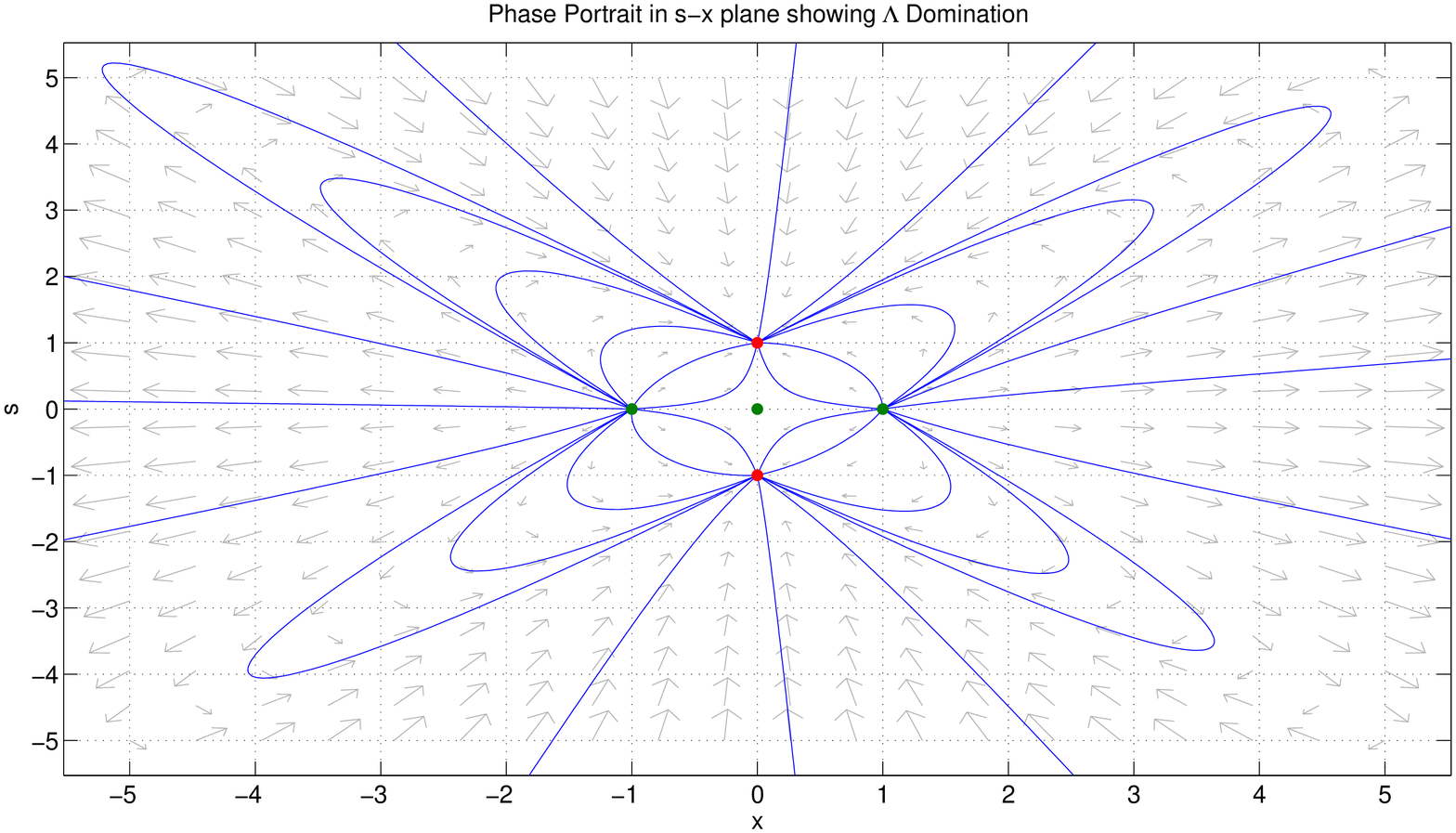}
\caption{Phase portrait in $s-x$ plane which shows that all trajectories eventually end in $\Lambda$ domination. The red points are equilibrium points and the green are unstable fixed points.}
\label{lambdaphase}
\end{center}
\end{figure}

Three of these fixed points are familiar from \LCDM: uncoupled matter domination, radiation domination, and $\Lambda$ domination. As with \LCDM\, our stability analysis reveals that $\Lambda$ domination is an attractor, whilst the other two are saddles. This is to be expected from the cosmic no hair theorem~\cite{PhysRevD.15.2738}. In our generalised Brans--Dicke model, we generically have three additional fixed points.  The first of these is always present, and corresponds to an era in which the kinetic energy of the scalar dominates the cosmic evolution. This fixed point is unstable. Whether or not the system evolves to this fixed point depends on the choice of initial conditions. For $|\alpha|< \sqrt{3/2}$ there exists a saddle point
corresponding to an era in which the coupled DM component and the kinetic energy of the scalar scale with one another, dominating the expansion. For $|\alpha|>\sqrt{1/2}$ there is another saddle point, corresponding to a three way scaling between coupled DM, radiation, and the kinetic energy of the scalar. These properties are explicitly demonstrated in the phase--space plot shown in Fig. \ref{lambdaphase}, where the evolution inevitably points towards the unique late time attractor corresponding to $\Lambda$ domination.

\section{Perturbation evolution}
\label{pertsec}

Now that we have studied the background evolution of our system, we can move on to the evolution of perturbations and hence examine how the coupling affects structure formation and the CMB anisotropy spectrum. We will work in the synchronous gauge and to first order (see e.g.~Refs.~\cite{Hwang:2001fb,Ma:1995ey}) where the line element is
\begin{equation}
{\rm d}s^2 = a(\tau) \{ - {\rm d}\tau^2 + (\delta_{ij} + h_{ij} ) {\rm d}x^i {\rm d}x^j \} \, ,
\end{equation}
\noindent and the metric perturbation $h_{ij}$ can be decomposed into a scalar, vector and tensor part. Note that we have switched to conformal time $\tau$ for convenience. Focussing on scalar perturbations, in Fourier space $k$ we set 
\begin{equation}
h_{ij}=\frac{1}{3} h \delta_{ij}-\left(k_i k_j -\frac{1}{3} \delta_{ij} k^2\right) \mu \,,
\end{equation}
and can use the following definitions
\begin{align}
\delta_n &\equiv \frac{ \delta\rho_n }{ \rho_n} \,, &
\theta_n &\equiv ik_j v^{j}_{(n)} \, ,
\end{align}
\noindent where each species $n$ has coordinate velocity $v_{(n)}{}_i$, density contrast $\delta_n$ and velocity field gradient $\theta_n$. We write the perturbed energy--momentum tensor for uncoupled CDM as
\begin{equation}
\delta T^0_{(c)}{}_0=-\delta \rho_c, \qquad \delta T^0_{(c)}{}_i=\rho_c v_{(c)}{}_i=-\delta T^i_{(c)}{}_0, \qquad \delta T^i_{(c)}{}_j =0 \, ,
\end{equation}
where the uncoupled CDM pressure and anisotropic stress have been set to zero. Conservation of the CDM energy--momentum tensor implies
\begin{equation}
\delta'_c = - \half h' \, ,
\end{equation}
where here, and subsequently, primes denote differentiation with respect to conformal time. As is well known, there is a residual gauge freedom which allows us to consistently set $\theta_{\rm c}=0$. The perturbations of the Standard Model components are treated in the usual way by evolving the full Boltzmann hierarchy in CAMB~\cite{Lewis:1999bs}.

The coupled dark matter species undergo a non--standard evolution.  Assuming that the perturbed contribution to the energy--momentum tensor in Einstein frame is given by
\begin{equation}
\delta T^0_{(q)}{}_0=-\delta \rho_q, \qquad \delta T^0_{(q)}{}_i=\rho_q v_{(q)}{}_i=-\delta T^i_{(q)}{}_0, \qquad \delta T^i_{(q)}{}_j =0 \,,
\end{equation}
we can use $\nabla_\mu  T_{(q)\nu}^\mu=\alpha T_{(q)\mu}^\mu \nabla_\nu \phi $ to demonstrate that
\begin{align}
\delta'_q &= -\theta_q - \half h' + \alpha \delta\phi'  \,, \\
\theta'_q &= -\theta_q \mathcal{H} + \alpha(k^2\delta\phi - \phi'\theta_q) \, ,
\label{qcpert}
\end{align}
where $\delta\phi$ is the perturbation in the scalar field and ${\cal H}=a'/a$. 

Finally, the evolution of perturbations in the scalar field is governed by the perturbed scalar equation of motion, and is given by
\begin{equation}
\delta\phi'' + 2\mathcal{H}\delta\phi' + k^2\delta\phi +\half h'\phi' = -\alpha a^2 \delta \rho_q  \, .
\end{equation}

We now modify CAMB~\cite{CAMB}, assuming adiabatic initial conditions, to incorporate perturbations in these new species along with the background equations, Eqs. \eqref{Friedmann} and \eqref{KGeqn}, to calculate the CMB anisotropy spectra. As discussed in Sec. \ref{backgroundsec}, we must evolve our background first, before evolving the perturbations. To do this, an adapted version of the quintessence module available for CAMB~\cite{CAMBquintmod} was used since this was designed to evolve quintessence models, which also do not have a \LCDM\ background.

\begin{figure}
\begin{center}
\includegraphics[width=1\textwidth,height=90mm]{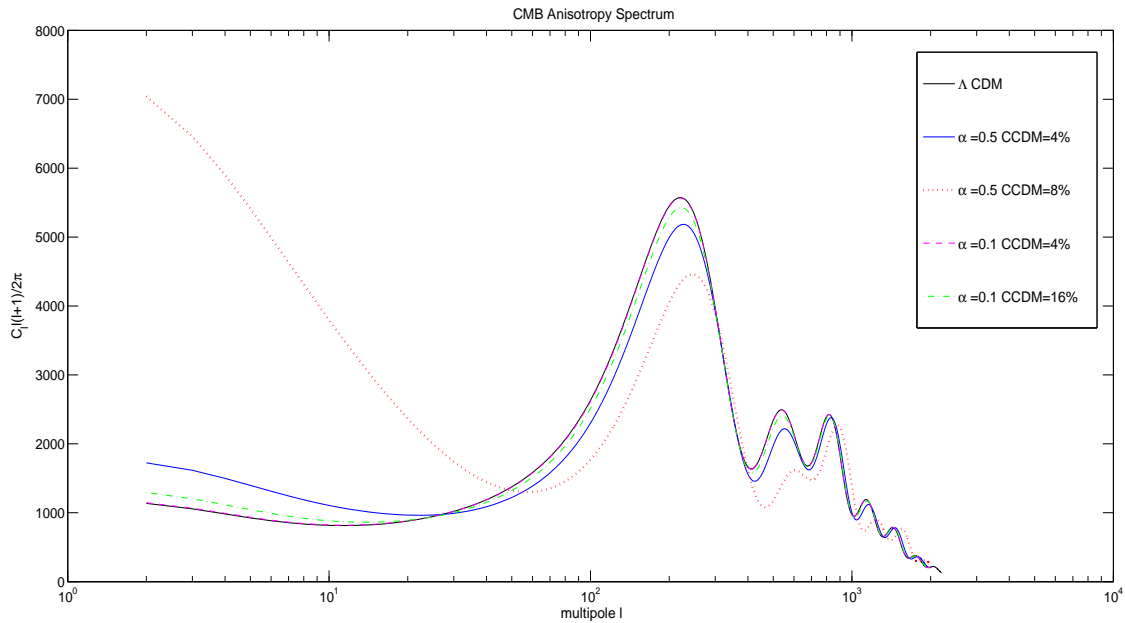}
\caption{CMB anisotropy spectra for various coupling strengths and proportions of coupled cold dark matter, CCDM.}
\label{CMB}
\end{center}
\end{figure}

In Fig.~\ref{CMB} we plot the resulting CMB anisotropy power spectra for various proportions of charged dark matter and coupling strengths, along with the spectrum for \LCDM. The most interesting thing to note from this plot is the fact that even at a level of 16\% coupled CDM  with coupling strength $\alpha=0.1$,  we still obtain a CMB power spectrum that is extremely close to \LCDM. As both the coupling strength and proportion of CDM coupled increase, we see greater deviation from \LCDM, to the point that a coupling strength of $\alpha=0.5$ or above is unlikely. 
These deviations are characterised by the large increase in power due to the Integrated Sachs--Wolfe (ISW) effect at low multipoles, and the lowering of the CMB peaks. The former has also been seen in similar models~\cite{Koivisto:2005nr,Koivisto:2004ne} and is mainly due to the late time variation of the gravitational potentials, the wells growing deeper in time as the scalar evolves to larger values. The latter is due to the fact that there is an increase in the coupled CDM density at early times, as is  evident from Fig. \ref{background}. This results in fewer baryons at the time of recombination, which in turn raises the sound speed and suppresses the amplitude of oscillations. Furthermore, if  matter--radiation equality is earlier then the potential at the peak is deeper due to less pressure damping resulting in a large redshift as the photons climb out after last scattering, again, implying a smaller final anisotropy (for an excellent discussion of these effects, see Ref.~\cite{Hu:2001bc}).  Note that the peaks are also shifted slightly to the right as we deviate more from \LCDM. This is also due to matter--radiation equality occurring at earlier times, such that only very small scales enter the sound horizon during radiation domination, therefore the power spectrum turnover occurs on smaller scales, also shifting the CMB peaks to smaller scales i.e. higher $l$. 

\begin{figure}
\begin{center}
\includegraphics[width=1\textwidth,height=90mm]{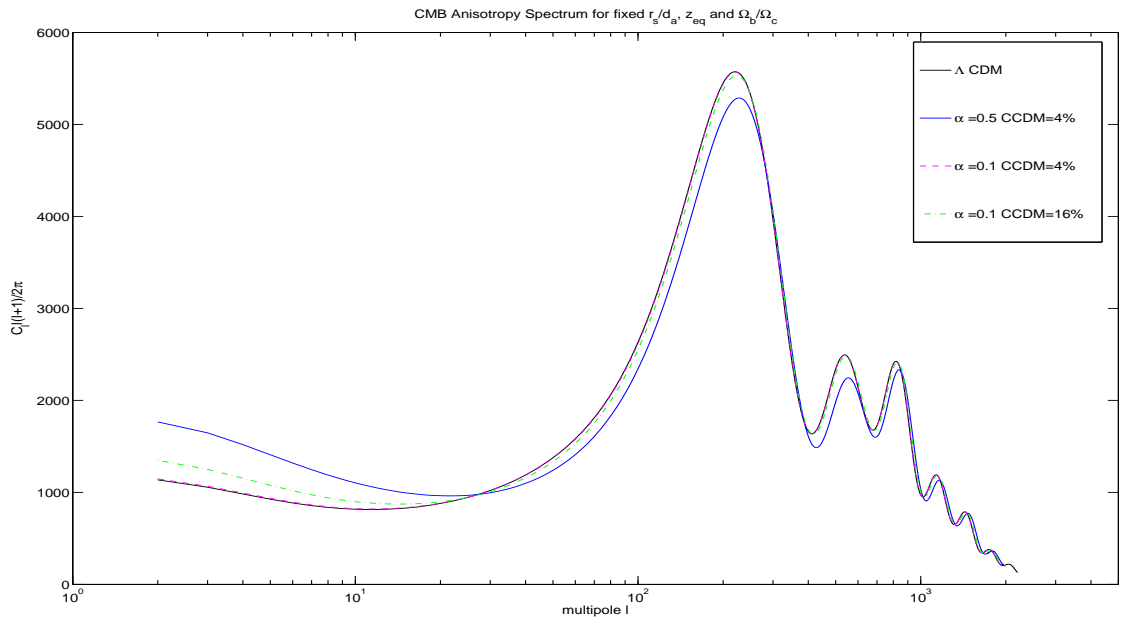}
\caption{CMB anisotropy spectra for various coupling strengths and proportions of coupled cold dark matter. In this case we have varied the baryon, total CDM and dark energy present day density parameters so that the angular size of the sound horizon at decoupling, the redshift of matter-radiation equality and the baryon fraction are fixed.}
\label{CMB_fixed}
\end{center}
\end{figure}

For the CMB anisotropy power spectra shown in Fig.~\ref{CMB}, we adopted the \LCDM \, best--fit values for $\Omega^0_{\rm b}$, $\Omega^0_{\rm c_{\rm tot}}$, and $\Omega^0_{\rm DE}$, and chose different combinations of $\Omega_{\rm q}$ and $\alpha$. Whilst this facilitates a direct comparison to \LCDM, this choice is not optimal for addressing the question of whether the new effects induced by the scalar field and coupled dark matter can be constrained by the CMB. In order to better isolate the effects on the CMB that are purely from the coupled CDM, we fix the following quantities to the values derived from the \LCDM\ fit to the WMAP data~\cite{wmapwebsite}: the redshift of matter--radiation equality, $z_{\rm eq}=3196\pm 74$, the ratio of baryons to CDM, $\Omega_{\rm b}/\Omega_{\rm c_{\rm tot}}=0.201 \pm 0.045$, and the angular size of the sound horizon at decoupling, $\theta_{\star} = 0.010391 \pm 0.000022$ (see e.g. Ref.~\cite{Dodelson:2003ft} for the definition of $\theta_{\star}$). This is done by varying $\Omega^0_{\rm b}$, $\Omega^0_{\rm c_{\rm tot}}$, and $\Omega^0_{\rm DE}$ for each combination of $\Omega_q$ and $\alpha$. In Fig.~\ref{CMB_fixed} we plot the CMB power spectra found using this method. We are able to achieve a better match to the observed peak structure for a coupling strength of $\alpha=0.1$. Even for 16\% coupled CDM the agreement is very good. However, when $\alpha$ becomes larger the match to the peak structure is barely improved by varying the density parameters. We have not included the power spectrum for 8\% CCDM and $\alpha=0.5$ in Fig.~\ref{CMB_fixed}, as in this case it was not possible to obtain a reasonable value for $z_{\rm eq}$. For a fixed present day matter density, equality occurs earlier if the fraction of coupled CDM is increased. The redshift of equality can be reduced, while keeping the fraction of coupled CDM today the same, by lowering the total matter density. In doing so  the kinetic energy of the scalar field is increased. As the scalar field grows, the large coupling leads to a significant enhancement of $\Omega_q$  via the conformal factor, $e^{\alpha \phi}$. This in turn causes the total matter density to increase. For 8\% CCDM and $\alpha=0.5$ this increase is so large that there is no combination of other parameters that reduces $z_{\rm eq}$ to the required value. For 4\% CCDM this was also a problem and a compromise had to be made between keeping the fixed quantities within their error bars and keeping the matter density consistent with other astronomical measurements, hence the minimal improvement in fit. It is interesting to note that this exercise also allows over half the dark matter to be coupled to the scalar field, provided the coupling is not too large.

\section{Conclusion} \label{sec:conc}

In this paper we have investigated the effects of coupling a variable fraction of the dark matter to a scalar field without the presence of a potential. We find that the background evolution can deviate significantly from that  of \LCDM, depending on the strength of the coupling and the proportion of CDM coupled. This means that when carrying out calculations that depend upon the background, such as perturbation evolution, it is necessary to take the evolution of the scalar field background into account.  This is due to the fact that we do not include a bare potential term, and so the effective potential does not have a local minimum at finite values of the scalar. For this reason, we also find that the late time attractor solutions often seen in quintessence models are not present here. In fact, the only stable solution to this model is $\Lambda$ domination, as one might have anticipated from the cosmic no hair theorem.

The main purpose of our work, however, was to  derive the perturbation equations and solve them in order to calculate the CMB angular power spectrum. Here we find that there is only a small deviation from the \LCDM\ predictions, even for  $\mathcal{O}(0.1)$ scalar coupling to matter, in gravitational units, with over half the total dark matter coupled directly to the scalar field.  This is quite compelling and certainly a surprising result. It is, in part, down to the fact that the coupling itself seems to play only a small role in the anisotropies. It is actually the relative proportions of matter and radiation present, and the time of matter--radiation equality, that has the biggest effect on the CMB. The key point is that  the CMB is only an indirect probe of long range dark matter interactions. CMB photons  do not couple directly to the scalar field, and only feel its effect through their impact on structure formation.  The scalar mediates an additional attractive force on the relevant dark matter species enhancing the growth of density perturbations, the effect being more pronounced the greater the coupling and the greater the fraction of coupled dark matter.

It would be interesting to study the effect on alternative probes of the dark matter from supernovae to large scale structure, and to jointly analyse these data sets along with CMB data. As we have already mentioned, supernovae might be considered as a more direct probe of the long range force as it causes galaxies, dominated by dark matter, to follow non-geodesic trajectories.  The additional attraction mediated by the scalar  is also expected to increase the number of large clusters, which is not favoured by observation, so a detailed studied of structure formation in this model may also place a strong bound on our parameters. For related work with coupled quintessence scenarios and chameleons see e.g. Refs.~\cite{tarrant2011} and~\cite{Brax:2013mua}. Furthermore, generalisations of the model considered here, including other non-trivial couplings between the scalar and certain species in the dark sector could also be studied. For example, one might consider a dark sector version of the BSBM model \cite{Sandvik:2001rv,Barrow:2011kr} or independent couplings to more than one matter species.

Finally, we conclude our discussion with a word about additional long range {\it repulsive} forces. Typically these require a vector field to mediate the interaction, but we might have hoped to mimic its effect by reversing the sign of the kinetic term in our scalar field model, bearing in mind that such a toy model should not be taken too seriously because of violations of unitarity. However, it turns out that this trick doesn't work. The ghost--like nature of the scalar turns the effective potential into an unstable one, leading to rapid evolution of the scalar and time varying gravitational potentials. This has a dominant effect on the CMB angular power spectrum at low multipoles through a large late time Sachs-Wolfe effect that would not necessarily be expected in a consistent scenario involving a vector mediator.

\subsection*{Acknowledgments}

We would like to thank Adam Christopherson, Nemanja Kaloper and Costas Skordis for useful discussions. 
SCFM and AMG are supported by STFC, AP by a Royal Society URF, and
ERMT by the University of Nottingham.

\bibliographystyle{ieeetr}
\bibliography{bibV13}

\end{document}